\documentclass[conference]{IEEEtran}
\usepackage{times,amsmath,epsfig,amssymb,cite,bm,float,epstopdf,graphicx,graphics,amsthm}
\usepackage{balance,multicol}
\usepackage{color}
\usepackage[stable]{footmisc}
\usepackage{amsthm}
\usepackage{pdfsync}
\usepackage{subcaption}
\usepackage{amssymb}
\usepackage{amsmath}
\usepackage{ifthen}
\usepackage{cool,booktabs}
\usepackage{makeidx}
\usepackage{verbatim}
\usepackage[ruled,vlined]{algorithm2e}
\usepackage[font=small]{caption}

\title{Machine Learning assisted Handover and Resource Management for Cellular Connected Drones}
\author{Amin Azari$^*$, Fayezeh Ghavimi$^+$, Mustafa Ozger$^*$, Riku Jantti$^+$, and Cicek Cavdar$^*$\\
$^*$KTH Royal Institute of Technology, Sweden $^+$Aalto University, Finland \\
Email: \{aazari, ozger, cavdar\}@kth.se, \{fayezeh.ghavimi, riku.jantti\}@aalto.fi}

\begin{document}
\maketitle

\begin{abstract}
 
Enabling cellular connectivity for drones introduces a wide set of challenges and opportunities. Communication of cellular-connected drones is influenced by 3-dimensional mobility and line-of-sight channel characteristics which results in higher number of handovers with increasing altitude. Our cell planning simulations in coexistence of aerial and terrestrial users indicate that the severe interference from drones to base stations is a major challenge for uplink communications of terrestrial users. Here, we first present the major challenges in co-existence of terrestrial and drone communications by considering real geographical network data for Stockholm. Then, we derive analytical models for the key performance indicators (KPIs), including communications delay and interference over cellular networks, and formulate the handover and radio resource management (H-RRM) optimization problem. Afterwards, we transform this problem into a machine learning problem, and propose a deep reinforcement learning solution to solve H-RRM problem. Finally, using simulation results, we present how the speed and altitude of drones, and the tolerable level of interference, shape the optimal H-RRM policy in the network. Especially, the heat-maps of handover decisions in different drone's altitudes/speeds have been presented, which promote a revision of the legacy handover schemes and redefining the boundaries of cells in the sky.

\end{abstract}

\begin{IEEEkeywords}
drone communications, handover, radio resource management, deep machine learning.
\end{IEEEkeywords}

\section{Introduction}
\begin{figure}[!tb]
    \centering
    \includegraphics[width=3.3in, height=1.5in]{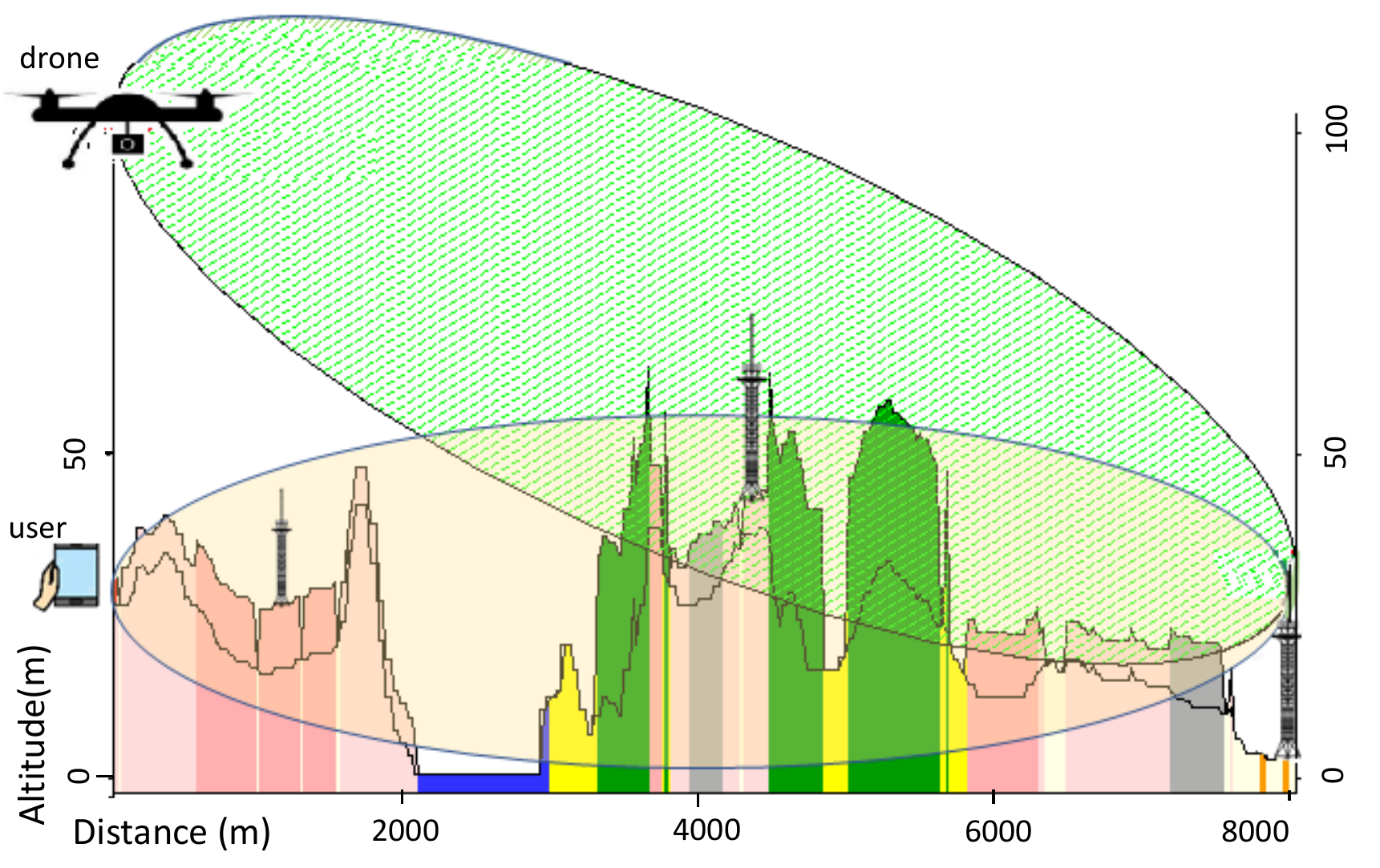}
     \caption{ Fresnel zones (elliptical region determined by distance between transmitter and receiver and operation frequency)  of a terrestrial and  drone user in communications with a far-away BS using Mentum Cell Planner.}
    \label{fig:f2}
\end{figure}

Commercial drone applications have attracted profound interest in recent years in a wide set of use-cases, including area monitoring, surveillance, and delivery \cite{SHayat16_survey}. In many applications, drones, also known as unmanned aerial vehicles (UAVs), require connectivity to carry their tasks out. Due to the ubiquitous coverage of cellular networks, they serve  as the major infrastructure for providing wide-area yet reliable and secure drone connectivity beyond visual line-of-sight range \cite{YZheng19_UAV_UE, SSekander18_multitier}. A comprehensive set of empirical analyses on providing connectivity for drones through LTE networks have been conducted recently in the context of 3GPP, and some of test-field results could be found in  \cite{3gpp_lte}. Regarding the fact that the probability of experiencing line-of-sight (LoS) propagation to the neighbor BSs increases with altitude \cite{int-uav}, the wireless channels between flying users and  neighboring base stations (BSs) experience almost free-space fading  \cite{3gpp_lte}. Hence, in the uplink direction,  drone  communications are expected to incur significant interference to uplink communications of terrestrial UEs, and in the downlink communications, drones are vulnerable to receive strong interference from neighbour BSs, as shown in Fig. \ref{fig:f2} \cite{HCNguyen18_reliability}. Furthermore, due to the mobility of drones in a 3-dimensional (3D) space without predetermined roads as compared to the legacy terrestrial UEs, radio resource provisioning in the sky becomes a difficult task in this dynamic environment in comparison to the legacy urban/rural service areas with predetermined buildings and roads.  Moreover, the terrestrial users usually receive strong signals from a few neighbor BSs. This makes the user-BS association problem less complicated in comparison with the drone communications in which, drones observe LoS signals from several BSs. By considering the speed of drones and the large-set of potential BSs that a drone can be served with them,  many handover events will be triggered for drones \cite{handover_ch}. Then, we observe that introduction of aerial users to cellular networks needs a revision of some communications protocols which have been developed by considering the legacy terrestrial users. Here, we focus on the handover and radio resource  management (H-RRM) problem in serving drone communications over cellular networks. In this problem, the key performance indicators of our interest are: (i) the interference of drone communications on terrestrial communications, and (ii) the experienced delay in drone communications.  Among candidate enablers for solving such a complex and dynamic problem, we leverage  machine learning (ML) tools, transform the problem into a  machine learning problem, and provide a solution for it. The proposed ML schemes enable cellular networks to capture the temporal and spatial correlations between decisions taken in the network in serving drones in order to make a foresightful and cognitive decision in later decision epochs.  To the best of authors' knowledge, our work is the first  in literature that investigates the  H-RRM problem in a network consisting of   drone and terrestrial users, and leverages a ML-powered solution for solving the problem. The key contributions of this work include:
\begin{itemize}
\item Analyze the received   interference from drone communications in the ground BSs using cell planning tools and  leveraging real geographical and land-use data.
\item
Formulate the H-RRM problem in serving drone and terrestrial users as a  machine learning problem by incorporating delay in drone communications and interference to terrestrial users in the design of the reward function.
\item Present a  reinforcement learning solution to the problem.
\item Present the impacts of  different system parameters  on the  H-RRM problem decisions. Analyze the interplay between level of  interference to terrestrial users, handover overhead, allocated resources to drones, and experienced delay. 
\item Present boundaries of cells  in the sky, handover regions, as a function of altitude and speed of drones, and level of tolerable interference to the terrestrial users.

\end{itemize}

The remainder of this paper is outlined as follows. Section II presents the  challenges, existing solutions, and research gaps. System model and problem formulation  are presented in Section III. Section IV presents the proposed solution. Simulation results are presented in Section V, followed by the conclusion given in Section VI.

\section{Challenges in Serving Aerial Users and State-of-the-art Solutions}\label{Chall}

\begin{figure*}[!t]
    \centering
    \includegraphics[width=0.95\textwidth]{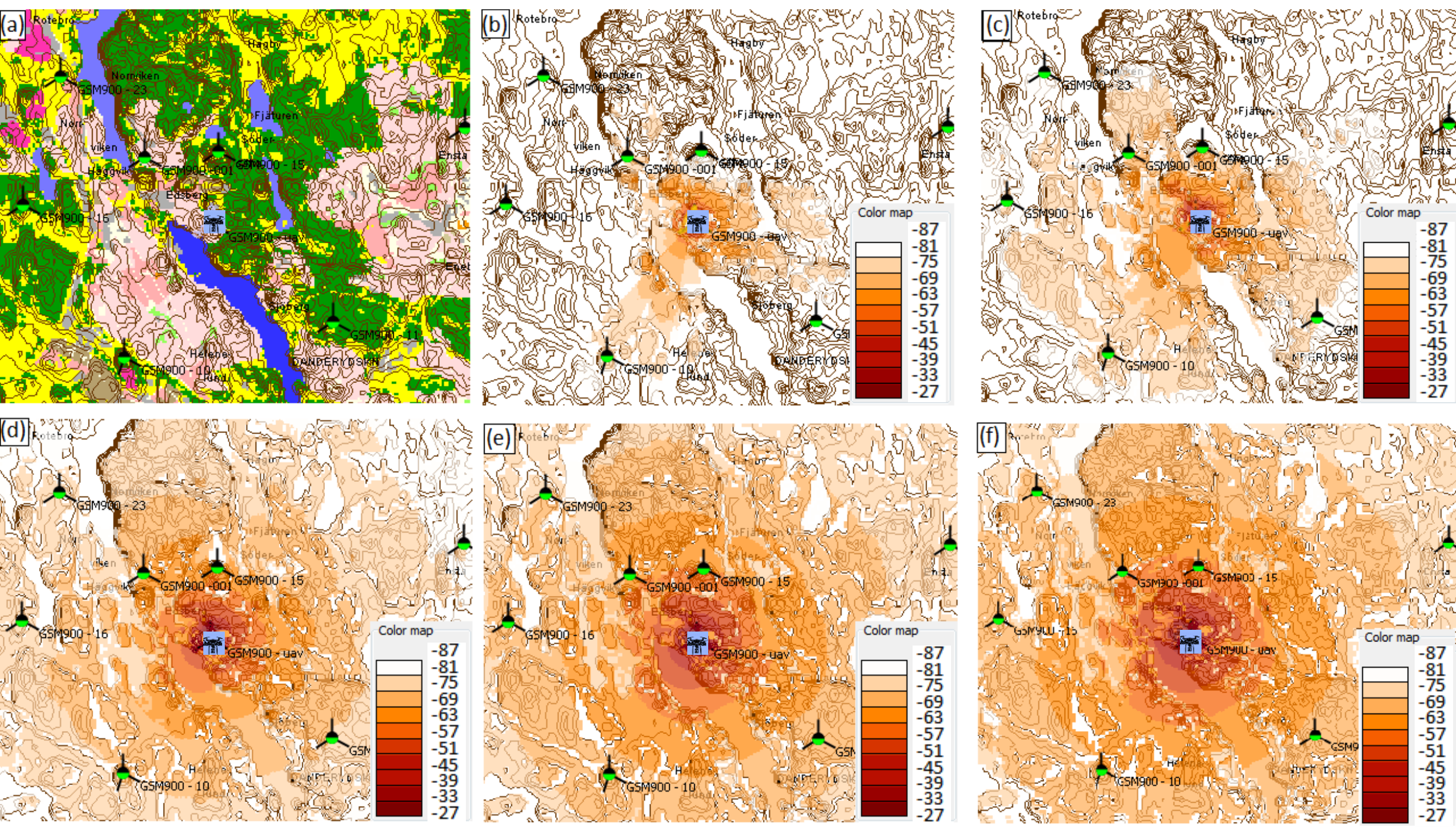}
     \caption{Assessment of interference (in dBm) changing with altitude of drone using Mentum Cell Planner. (a)  Partial map of the Stockholm; (b)-(f) Received interference  from a drone at altitude $h$ at BSs (tower height: 25m). Drone transmit power is 23 dBm  and  Tx/Rx antenna gain is 0dB. $h$=\{1.5m, 20m, 70m, 150m, 200m\} for (b)-(f) scenarios respectively.}
    \label{fig:f1}
\end{figure*}

\noindent \textbf{Interference on Terrestrial Communications.}  Connectivity for aerial users over cellular networks should be enabled in a way to minimize the side-effects on quality-of-service (QoS) of terrestrial users. Unfortunately,  the favorable propagation conditions that drones enjoy due to their LoS link to the associated base station degrades QoS of terrestrial users. Fig. \ref{fig:f1} represents our simulations results using Mentum Cell Planner tool\footnote{developed by Ericsson, available at https://www.infovista.com} leveraging real geographical and network data for Stockholm, Sweden. This figure shows that the received interference from a drone increases significantly, and extends to a much wider area, by increasing the altitude of the flying drone. In \cite{int-uav}, the impact of air-to-ground communications on the  coverage performance of   cellular networks  has been investigated. In \cite{path},  a deep reinforcement learning path planning for drones in investigated, in which, each drone aims at making a balance between its energy efficiency and  interference to the ground networks along its path. The authors in \cite{icic} maximize the weighted sum-rate of the drone-to-BS link and existing ground users by jointly optimizing the drone's uplink cell associations and power allocations over multiple shared radio resource
blocks. 

\noindent \textbf{Interference on Drone Communications.} 
In \cite{int-uav}, it has been shown that the LoS propagation conditions a drone user experiences at high altitudes actually result in an overall negative effect, as the high vulnerability to interference from BSs in the downlink direction dominates over the increased received signal power from the associated BS. From Fig. \ref{fig:f1}, which  represents the received interference from drone to BSs in the uplink direction, the performance in the downlink direction for drone will be much severe because drone will receive many interfering signals from neighbor BSs, where the BSs have a much higher transmit power in comparison with the drones. 


\noindent \textbf{Handover for Drone Communications.} 
Current handover mechanisms for terrestrial UEs mainly trigger a handover based on a policy which is applied on a set of metrics, such as received signal received quality (RSRQ).  Regarding the LoS links between drone and neighbor BSs, the set of neighbor BSs that a drone can connect has a much higher cardinality as compared to a terrestrial user. The movement of the drone along its trajectory triggers frequent measurement reports, which result in frequent handovers in case the handover decision is made by the legacy scheme of comparing received signal strengths (RSSs). In \cite{handover_ch}, handover performance of LTE-Advanced networks when serving drones is investigated in a measurement campaign. It has been shown that a drone flying at $150$ m altitude has experienced $5$ times more number of handovers than a ground UE with the same speed. Then, the legacy handover management mechanisms lose their merits in serving flying UEs, triggering unnecessary handovers. 
In the reverse setup when drones are serving as BSs, \cite{QLi18_RL} solves the handover management problem by leveraging reinforcement learning. However, this is a very different problem since drones are stable and used as coverage BSs.
 
\noindent \textbf{Research Gap.} Terrestrial cellular networks will face new challenges serving drone users: 1) Unnecessary handovers due to the increasing LoS probability with the increase of altitude, 2) interference created by flying UEs over terrestrial networks blasting power over a large area. This motivates us to study uplink communications of a cellular network serving both flying and  terrestrial users, and aiming at making H-RRM decisions in a way to maximize QoS of drone users with minimum impact in terms of interference on the terrestrial users.

\section{System Model  and Problem Formulation}\label{SecInt}

We focus on the uplink (reverse link) of cellular mobile networks supporting drone and terrestrial UEs over a service area. The set of available BSs in the service area is denoted by $\mathcal L$.  A co-channel deployment is considered, in which BSs operate in a system with a bandwidth $\mathbb{B}$ consisting of $N_b$ radio resource blocks (RRBs). We focus on the  handover and resource management problem. Then, at the decision epoch $t$, based on the status of terrestrial users and network resources, we aim at deciding which BS with which set of resources should serve the drone, and what should be the transmit power. 

\subsection{Air-to-Ground Channel and Modeling of the KPIs}\label{kpim}

\noindent \textbf{The Air-to-Ground (A2G) channel.} The A2G channel depends strongly on the presence of LoS propagation characteristics between the drone and BS. Without loss of generality,  in the following we provide the A2G channel model for an urban environment, while extension to suburban and rural areas is straightforward by substituting the respective parameters using the channel conditions in \cite{3gpp_lte}. The probability of experiencing LoS propagation in  communications between drone, at altitude $h$ with speed $v$ with respect to the $l$th BS is given as \cite{3gpp_lte}: $
    \mathcal P_\text{LoS} = \ {d_1}/{d_\text{2D}^{l}} + \exp(\ {d_1}/{p_1})(1\text{-}\ {d_1}/{d_\text{2D}}),$  for $d_\text{2D}^{l}$$>$$d_1$, and 1 otherwise. In this expression, $   p_1=4300\log_{10}(h)-3800,$ $d_1=\max((460\log_{10}(h)-700), 18),\hspace{1mm}\text{ in meters}$, and  $d_\text{2D}^{l}$ is the horizontal distance between the drone and the BS $l$ for $22.5 <h \leq 100$ (in meters). For $100 < h \leq 300$, $\mathcal P_\text{LoS}=1$ is assumed. For LoS conditions, $\text{PL}_\text{LoS}$, when $22.5 <h \leq 100$, is given as \cite{3gpp_lte} \begin{equation}
    \text{PL}_\text{LoS} = 28 + 22 \log_{10}(d_\text{3D}) +20 \log_{10}(f_c),\label{pl3}
\end{equation}  where $d_\text{3D}^{l}$ is the 3D distance between the drone  and the $l$th BS  in meters, $h_\text{BS}$ is  the height of the BS, and $f_c$ is the carrier frequency in GHz. Path-loss for non LoS condition (NL), $\text{PL}_\text{NL}$, is given as \cite{3gpp_lte}  
\begin{equation}
    \text{PL}_\text{NL}  \text{=} 15 \text{+} ( 46\text{-}7\log_{10}(h) ) \log_{10}(d_\text{3D}) \text{+} 20 \log_{10} ( {f_c} ).
\end{equation}
Shadow fading also depends on the LoS/NL condition. Standard deviation of the shadow fading  for LoS condition is $\sigma_\text{LoS} = 4.64~\text{exp}(-0.00066h)$, and for NL condition, $\sigma_\text{NL} = 6$ \cite{3gpp_lte}. Finally, the channel is assumed to exhibit Rayleigh block fading characteristics. 

\noindent \textbf{Buffer Queue Size and Delay.}
First, we focus on the communication delay for the drone. Let $u(t)$ denote the number of data units in bits that arrive at the buffer of  drone  at the end of subframe $t$. The arrival of the data units follows a Poisson point process (PPP) distribution with associated parameter $\lambda^0$. One must note that the handover decision affects the arrival rate  as some control data is needed to be transferred as well. Here, we model the arrival rate of data units to the buffer queue of drone as follows $
\lambda(t) = \lambda^0 + \sum\nolimits_{\tau=1}^{T_h} I(t-\tau)\lambda_{h}^{m},$ 
in which $\lambda_{h}^{m}$ models the arrival rate of control signals due to handover, $T_h$ is the length of time at which control messages will be issued after a handover decision, and $ I(t-\tau)$ is a handover indicator function, equal to one if a handover has happened for drone at $t-\tau$, and zero otherwise. Then, the overall modeling of control and application data to the drone buffer will follow a Switched PPP (SPP), with light and heavy traffic arrival windows. Hence, the drone's buffer queue size as one of our KPIs is modeled as $ q(t+1) = q(t) + u(t) - s(t),$ where $q(t)$ indicates the number of data units in the buffer of the drone  at the beginning of subframe $t$, and $s(t)$ is the successful data units transmitted from the buffer of the drone  during the transmission interval. The expected queuing delay for a newly added packet to the buffer at time $t$ could be expressed as $q(t)/R(t)$ where $R(t)$ is the expected data rate for drone, and is affected by H-RRM  decisions. 
To capture the delay threshold as our KPI, we consider a maximum buffer  size, i.e. $q^{\text{max}}$, beyond which, a packet is  dropped. 
 
\noindent \textbf{Allocated Spectrum and Data Rate.}
By assuming  that the coherence time of the channel is considered to be greater than a transmission time interval (TTI), the achieved uplink data rate for drone over the allocated subset  of subcarriers, denoted by $\bf{b}$(t), is derived as:   
$R(t) = {\mathcal F}(W_s,{\bf{b}}(t), {\bf{p}}(t),{\bf h}(t)),$
in which, ${\mathcal F}(\cdot)$ is a function, to be described in the following. Furthermore, $W_{s}$ is the subcarrier bandwidth, $ {\bf{p}}(t)$ is the power allocation vector, ${\bf h}(t)$ is the vector of ratio of channel gains to the noise plus interference level over the allocated set of subcarriers. Without loss of generality, we exemplify our modeling for single carrier frequency division multiple access, and  approximate the ${\mathcal F}(\cdot)$ function  as $R(t) \approx W_s|{\bf b}|\log_2(1+\gamma(t)),$  where 
$
\gamma(t)  = P(t)(\sum_{s\in{\bf b}(t)} \frac{1}{\alpha_{s}(t)})^{-1},  \alpha_{s}(t) = \frac{|H_{s}(t)|^{2}}{N_{0}+W_sI_{s}(t)}.$
 $N_0$ is the noise power over each subcarrier,  $I_{s}(t)$ is the power density of interference over $s$th subcarrier, and $P(t)$ is the transmit power \cite{Kalil}.  Furthermore, ${\bf b}(t)$ is characterized by a RRB allocation indicator function, denoted by $\theta(l,b,t)$, which is 1 if the $b$th RRB of BS $l$ is allocated at time $t$ to the  drone, and 0 otherwise.

\noindent \textbf{Interference to Neighbor BSs.}
The interference incurred by the uplink transmission of the drone to $l$th neighbor BS, $\text{Int}(l)$, is calculated as $P - \text{PL}_\text{X}(h, d_\text{3D}^{l}) +G_{\text{tx}}+G_{\text{rx}}$ in dBm. In this expression,  $P$ is the drone transmit power, $G_{\text{tx}}$ is the transmit antenna gain (usually isotropic with zero gain), $G_{\text{rx}}$ is the receive antenna gain (depending on the altitude of drone and  radiation pattern of the receive antenna, $G_{\text{rx}}$ is determined), $X\in\{\text{LoS, NL}\}$ represent the LoS/NL condition, and $d_\text{3D}^{l}$ is 3D distance between the drone  and $l$th neighbor BS. Then, the  uplink interference from the drone to the  BSs in the service area could be  modeled as $ \sum\nolimits_{l\in\mathcal L }~\mathrm{Int}(l)$.


\subsection{Formulation of the H-RRM Optimization Problem}
Given status of drones at time $t$, $S(t)$, $\forall k\in\mathcal K$, as well as the available RRBs at each BS, i.e. $\mathcal B_l(t)$, $\forall l\in \mathcal L$, the problem is to find ${\mathbf{\theta}}(l,b,t)$ and ${P(t)}$, $\forall  b, l$; in order to satisfy the delay threshold of drones with a minimum amount of allocated radio resources, minimum  number of unnecessary handovers, and  minimum interference on ground BSs. Then, at decision epoch $t$, we need to solve the following optimization problem for serving the drone: 
\begin{align}
&\min_{\theta, P} \big[\frac{\xi_d q(t)}{R(t)}\text{+}{\xi_{\text s}}{\sum\limits_{l\in\mathcal L}\sum\limits_{b\in\mathcal B_l(t)}~\theta(l,b,t)}\text{+} {\xi_{\text f}}{  \sum\limits_{l\in\mathcal L}~\mathrm{Int}(l)}\text{+}\xi_{\text h} I(t)\big],\nonumber\\
&\textrm{subject }\textrm{to: }(\textbf{C1}) \sum\nolimits_{l\in\mathcal L}\sum\nolimits_{b\in\mathcal B_l(t)}P(t)\theta(l,b,t) \leq P^{\text{max}}; \nonumber\\
&(\textbf{C2})\text{ } q(t) \leq q^{\text{max}}; (\textbf{C3})\text{ } \sum\nolimits_{l\in \mathcal L}\mathcal I(0<\sum\nolimits_{b\in\mathcal B_l(t)}\theta(l,b,t))  {=} 1, \nonumber
\end{align}
in which, $\text{C}_1$ stands for the maximum allowable transmit power, $\text{C}_2$ stands for the delay threshold, and finally $\text{C}_3$ assures that the drone receives service from one cell only. Furthermore, $\mathcal I$ is an indicator function with binary output, and $\xi_x$, for $x\in\{\text{d,s,f,h}\}$, represents the scaling coefficient of  experienced delay, number of allocated RRBs, incurred interference, and  handover indicator, respectively. 

 \smallskip
\noindent \textbf{Solving the H-RRM Problem}
One observes that the H-RRM problem is not only a highly-complex non-convex optimization problem, but also the impact of decisions at time $t$, e.g. a handover, propagates in time and  affects different KPIs in later epochs. Then, we need to transform the problem to a problem in which, long-term benefits are taken into account along with the temporal QoS measures. Furthermore, the solution due to the dynamic properties of the cellular network environment, e.g. number of active BSs and users, making the solution adaptive to the changes in the environment is favorable. These requirements motivate us to transform this optimization problem into a reinforcement learning problem in which, the learning rate and forget factor tune the balance between temporal and long-term QoS measures.  In such a  problem,  the network's constraints are transformed to  the action and state spaces, and the objective function to be maximized is transformed to the reward function. In the following, we present the transformed problem and its solution.

\section{The  Transformed H-RRM Problem and  Solution}
In this section, we present the transformed H-RRM problem and propose an  algorithm to learn a policy, based on which, H-RRM decisions could be made based on the previous experiences.  Note that we assume the handover management and resource allocation decisions in the service area are taken by a central entity, hereafter called controller.  

\subsection{The  Transformed H-RRM Problem}
Reinforcement learning is a branch of ML dealing with an agent that aims at  taking \textbf{actions} in an  environment (described by \textbf{states}) so as to maximize its cumulative weighted \textbf{rewards}. Such a problem could be defined by definition of the states, actions, and the reward function, as follows.

\smallskip
\noindent \textbf{The State Space.} The state space describes the environment in which an agent is selecting its actions. We present the state of environment for the flying drone at time $t$ as $S(t)=[h(t), v(t), B(t),  q(t), \text{PL}_{l\in\mathcal L}(t)]$, consisting of its altitude, velocity, current serving BS, buffer queue size, and the last path-loss measurements to neighbor BSs, respectively. Hence, the state space $\mathcal S$ includes all potential realizations of $S(t)$. For example, $S(t)=[100, 20, 1,  0, \{80,90\}]$ represent a state in which, a drone at altitude 100 m  with 20 m/s speed is experiencing 80 and 90 dB path losses from BS-1 and BS-2, and is served by BS-1, and its buffer queue is empty. 

\smallskip
\noindent \textbf{The Action Space.} The action space presents the set of decision parameters available to the agent at each decision epoch. We present the action space at time $t$ as  $A(t)$=[$P(t)$, $\theta(l,b,t)$], where $P$ and $\theta$ stand for transmit power and radio resource allocation, as described in Section III. Then, the action space, $\mathcal A$, consists of different combinations of transmit power, associated BS, and allocated set of RRBs.   For example, $A(t) =[ 23 \text{ dBm}, \{2,0\}]$ represents an action in which, drone should transmit its data over two chunks of radio resources in BS-1 with 23 dBm transmit power. 

\smallskip
\noindent \textbf{Reward Function.} The reward function should mimic the objective function of the H-RRM problem to be maximized. By following the notation in the H-RRM problem in Section III, the immediate reward for serving drone  at time $t$, i.e. $w(t)$, is defined as a weighted sum of the rewards from resource efficiency in communications, low-delay performance, low-interference, and low number of unnecessary handovers. Then, we formulate  $w(t)$ as:
\begin{align}
  w(t)\text{=}&\alpha_{\text s}\times\text{Resource-efficiency Rew}\text{+}  \alpha_{\text d}\times\text{Low-delay Rew}\nonumber\\
  &\text{+}\alpha_{\text f}\times\text{Low-interference Rew}\text{-}\alpha_{\text h}\times\text{Handover Regret},\nonumber\\
 \text{=} &  \frac{\alpha_{\text s}}{1\text{+} \sum\limits_{l,b}\theta(l,b,t)} \text{+}  \frac{\alpha_{\text d}}{1\text{+}  q(t\text{+} 1)}\text{+}  \frac{\alpha_{\text f}}{1\text{+} \sum\limits_{l}\mathrm{Int}(l)}\text{-}\alpha_{\text h} I(t). \label{eq:rew}
\end{align}
In this expression, Rew is the abbreviations for reward, and $\alpha_{\text s}$, $\alpha_{\text d}$, $\alpha_{\text f}$ and $\alpha_{\text h}$ are the weights, which are determined from the relative importance of the KPIs in the target application. The first term  in $w(t)$ is proportional to the inverse of amount of consumed radio resources, and increases by a decrease in number of radio resources allocated to the drone. The second term is proportional to the inverse of buffer queue size, and the third term is proportional to the inverse of interference to the  BSs. Finally, the last term is an indicator of handover, and hence, represents the regret by a minus sign. Furthermore, all these measures have been scaled in [0,1] interval, in order to prevent one metric dominating others.

\subsection{Learning from Past Actions}
To leverage from past experiences in action selection, we use Q-learning as a model-free reinforcement learning algorithm since we need to learn a policy. 

\smallskip
\noindent \textbf{Q-learning and deep Q-learning.}  In Q-learning, action selection is done using an action-value function by following a policy. Each policy provides a mapping from the state space to the action space. The state-action value function  coupled with policy $\pi$, denoted by $Q_{\pi}(s,a)$,  is defined as the long-term expected accumulated discounted reward of state $s$ when action $a$ is taken, and the future actions are taken by following policy  $\pi$. In the basic Q-learning, each time an action is taken, its respective state-action value  is updated by 
\begin{align}
Q_{\pi}(s,a) {=} (1{-}\alpha)~Q_{\pi}(s,a) {+}\alpha[w~{+}\beta\max_{\mathbf{a\in \mathcal A}}~Q_{\pi}(s,a) ],   \label{eq:qupdate}
\end{align}
where $\beta$ is the discount factor and $\alpha$ is the learning rate ($0<\alpha<1$), which determines to what extent the learned Q-value will be updated. The convergence of $Q_{\pi}(s, a)$ to the optimized value  by adapting the learning rate has been proven in \cite{Watkins}. 
In the Q-learning, the Q-function is represented by a Q-table, containing states as rows, actions as columns, and values as entries. In a practical network, the number of states and actions could be so high that it is not possible to save the Q-values in a matrix. Then, either the states should be quantized for decreasing the dimensions of the Q-table, or the Q-function should be approximated by a neural network, i.e., deep Q-learning could be used \cite{mh}. We adapt deep Q-learning to solve the problem of handover and radio resource management (H-RRM) for flying UEs.

\begin{algorithm}[htb]
\nl Initialize: State space $S$, action space $A$, and the reward function coefficients, $C$, $\beta$, $\Phi$, and $\Phi_{\text old}$\;
\If{training iterations $<$  target}{
- Simulate the service area, and terrestrial and  drone user at random locations\;
\nl\If{drone inside service area}{
- Update state of drone, terrestrial users and network resources \;
- Take an action randomly from  action set\;
- Calculate the reward from \eqref{eq:rew}\;
- Save (current state:$s$, action: $a$, reward: $w$, and next state: $\bar s$) in memory\;
 (Experience replay) Take a mini-batch of memory, containing $M$ tuples of ($s, a, r, \bar s$)\;
\For{$m=1,2,\cdots$, $M$}{
 -\uIf{$\bar s$ is in the action space}{
   $y=w+\beta\max_{x\in A}Q(\bar s,x;\Phi_{\text{old}})$\;
  }
\textbf{ else }{     $y=w$\;
  }
  - Perform a gradient descent step on $(y-Q(s,a;\Phi))^2$ w.r.t. $\Phi$, and update $Q$\;
  \uIf{$\mod(m,C)$==0}{
   $\Phi_{\text{old}}=\Phi$\;}
}
}
}
\caption{Training  ML-powered H-RRM }\label{ho}
\end{algorithm}

\begin{algorithm}[htb]
\nl Initialize: As in Algorithm 1\;
 \nl \For{$t=1,2,\cdots$}{
 -\If{drone inside service area}{
- Update state of drone and remaining radio resources from terrestrial users\;
- Take action randomly with probability $\epsilon$, or optimally based on the  $Q(s,a;\Phi)$ with $1-\epsilon$:\\
\hspace{4mm}$\arg \max_{x\in  A, x:\text{ feasible at } t}\hspace{1mm} Q(s,x;\Phi)$\;
- \Return $P(t)$, $\theta(l,b,t))\hspace{1mm} \forall b,l$\;
 - Receive the reward $w$, calculate next state $\bar s$, and save ($s$, $a$, $w$, $\bar s$) in the memory\;}
  -\If{Need for policy update}{
  -Run Algorithm 1\;}
}
\caption{The ML-powered H-RRM }\label{ho}
\end{algorithm}
\subsection{The Solution: ML-powered H-RRM}
Given the above descriptions, here we present an algorithm for leveraging deep Q-learning in decision making for H-RRM of drone communications. The proposed algorithm is clarified in details as follows. First,  we zero-initialize the weights ($\Phi$) of a neural network used for approximating the Q-function. Then,  the neural network approximating the Q-function  is trained using a network simulator, as outlined in Algorithm 1. Then, at each decision epoch,  an action, including the serving BS, set of RRBs, and transmit power, is taken either randomly, with probability $\epsilon$, or greedy, with respect to the output of  the neural network with probability $1 - \epsilon$. Then, the received  reward is calculated from \eqref{eq:rew}. One must note that the update of the neural network  does not happen based on observations at each decision epoch. Instead, we save the (state, action, reward, next state) data in memory, and leverage memory replay and gradient descent for updating the neural network, as has been outlined in \cite{mh}. Finally, when there is need in the policy update, Algorithm 1 could be run in order to update the past learned weights using recent observations.  Algorithm 2, represents the details of action selection in the proposed solution.


\begin{table}[t!]
 \centering \caption{Parameters for performance analysis.}\label{part}
 \footnotesize
\begin{tabular}{p{4.3cm}p{3.72 cm}}\\
\toprule[0.5mm]
\textbf{Parameters} & \textbf{Values}\\
\midrule[0.3mm]
Service area & $500\times500 \text{ m}^2$ \\
BSs' positions& $(50,100);(200,400);(450,50)$\\
Available RRBs for drone (per TTI)&  Random, up to 4 $\times$ 180 KHz\\
BSs antenna height, carrier frequency& $25$ m, 2 GHz\\
Packet arrival rate and size at drone& $0.3$ Hz; $2$ Kbits \\
Handover control packet size &  4$\times$ 1Kbits \\
Circuit power, transmit power & $0.05$, $0.2$ Watt\\
Learning rate, discount factor& $1/(1+n/5)$, $0.8$\\
Minimum interval between handovers& $10\times$TTIs, and  TTI=0.001 sec\\
Drone speed and height& Default: $15$ m/sec, $50$ meters\\
\bottomrule[0.5mm]
\end{tabular}
\end{table}

\section{Performance Evaluation}

\smallskip
\noindent \textbf{Simulation Setup} We consider a service area of $500\times500$ m$^2$, in which $3$ macro BSs are  observable to  drones, and can simultaneously serve terrestrial and drone UEs. The simulator has been developed in Matlab, and implements Algorithm \ref{ho} in the Q-table form. In the benchmark scheme, RSS is used for handover management, i.e., if the received power from the target BS is $7$ dB stronger than the serving BS, a handover is triggered.   Furthermore,  in the benchmark scheme, all the available RRBs in the serving BS are allocated to the connected drone, while in the learning scheme, the allocation is determined by the policy from the Q-function. For the following analyses, we assume a drone at a constant altitude $h$ and speed $v$ is crossing the service area, and our  aim  is to (i) associate it to the best serving BS(s) at each radio frame, i.e., $10$ TTIs; and (ii) to allocate a set of  RRBs  at each TTI.  Learning rate is defined as $1/(1+n/5)$, where $n$ is the number of visits to the states, to increase the convergence rate.  The other system parameters used for the simulation are in accordance with \cite{3gpp_lte}, and could be found in Table \ref{part}. 


\begin{figure}[t!]
        \centering
         \begin{subfigure}[t]{0.45\textwidth}
        \centering
               \includegraphics[width=1\textwidth]{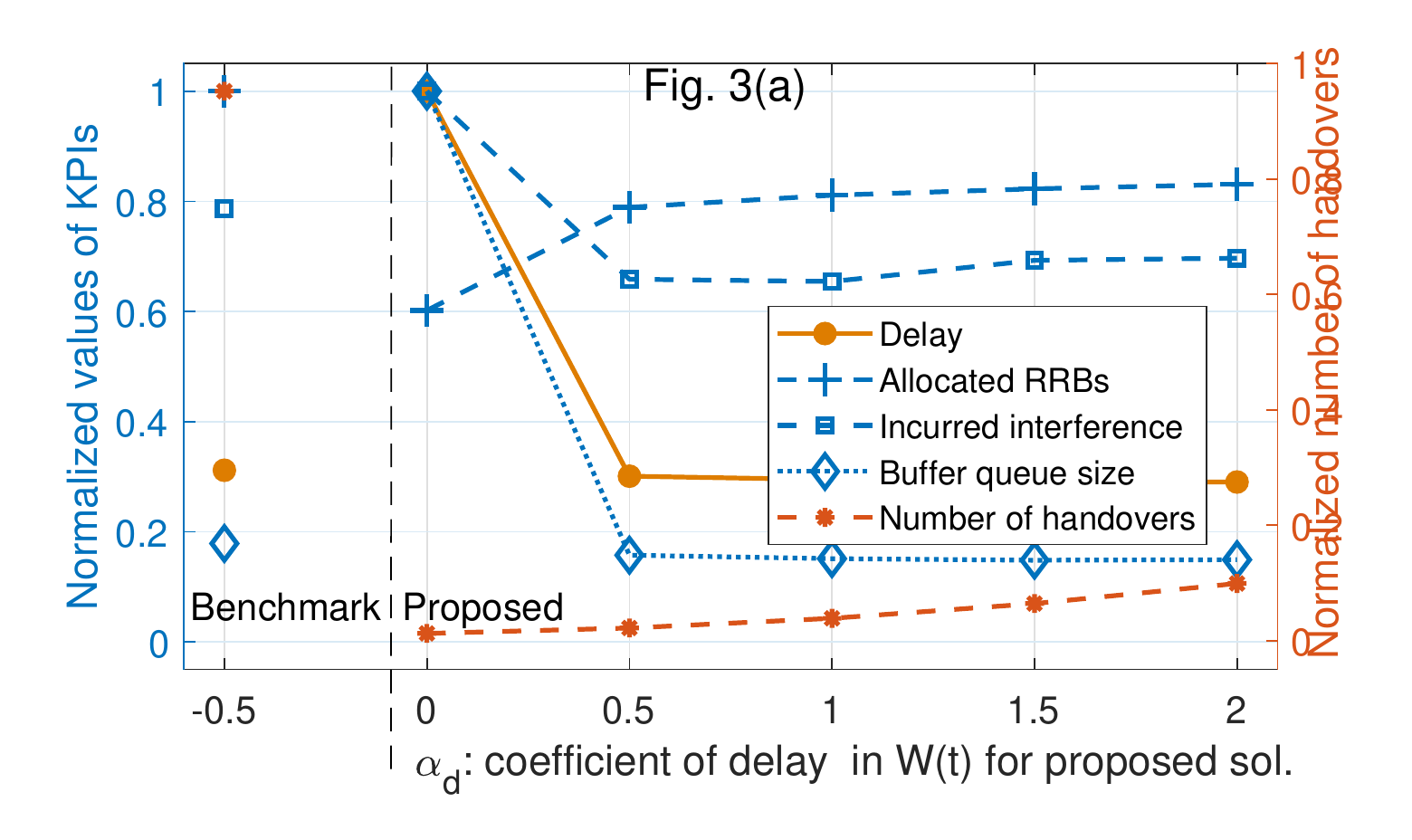}
\end{subfigure}\\
\begin{subfigure}[t]{0.45\textwidth}
        \centering
               \includegraphics[width=1\textwidth]{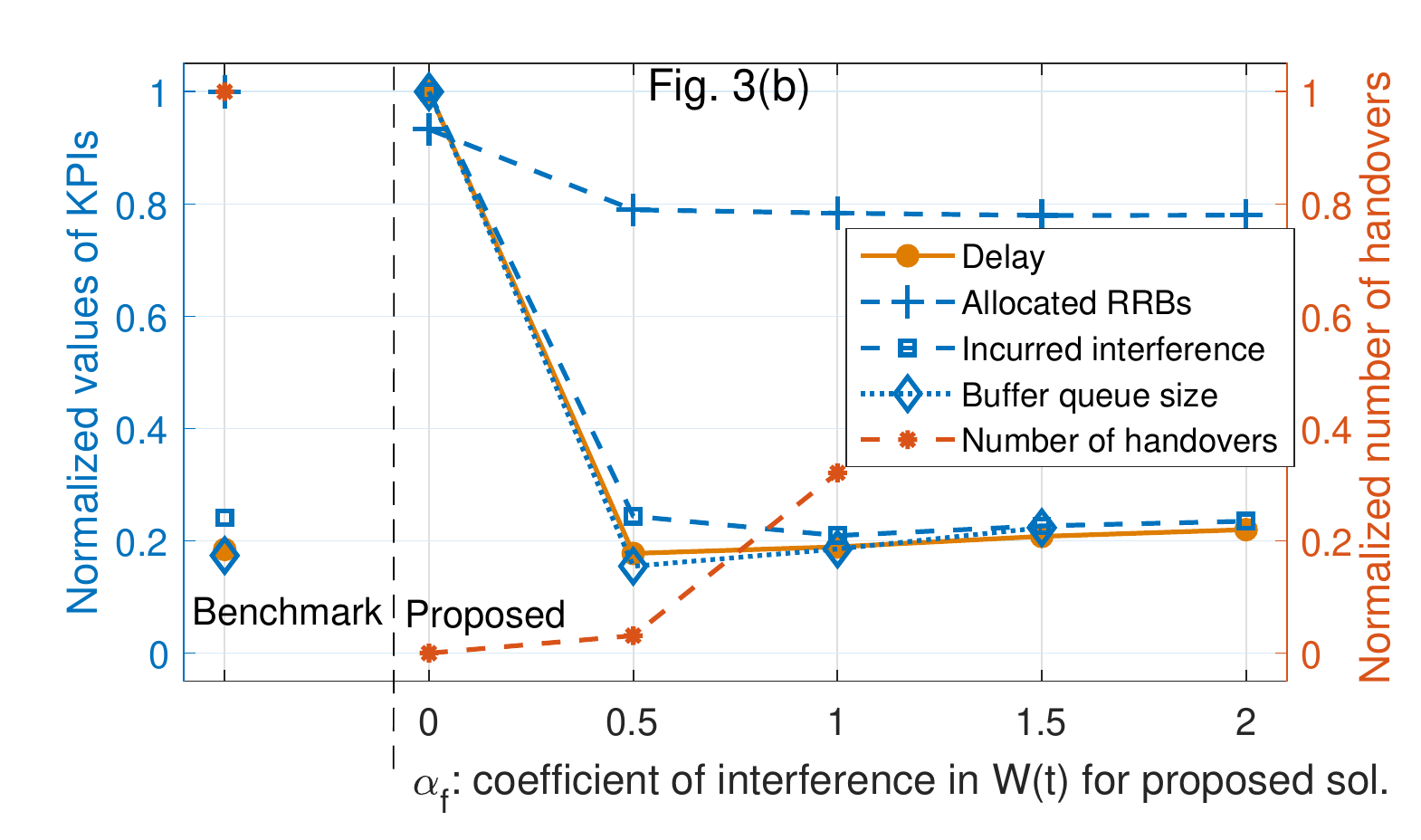}
\end{subfigure}\\
\begin{subfigure}[t]{0.45\textwidth}
        \centering
              \includegraphics[width=1\textwidth]{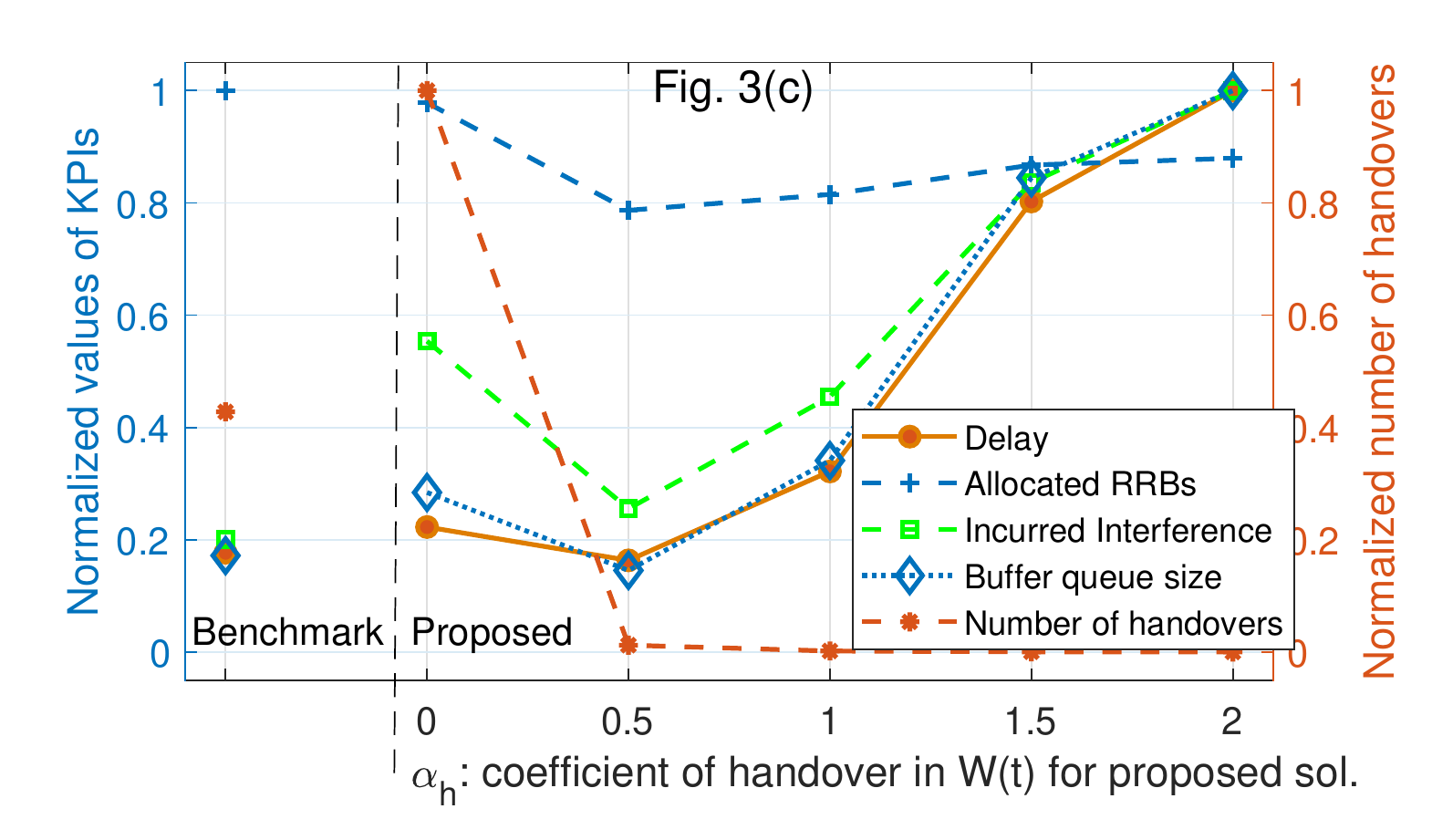}
\end{subfigure}

\caption{Performance impacts of changing coefficients of delay, interference, and handover. }\label{figam}
\end{figure}

\smallskip
\noindent \textbf{Simulation Results} 
Fig. \ref{figam}(a)-\ref{figam}(c)  evaluate  dependency of  KPIs of interest on their respective coefficients  in the reward function formulation, as described in \eqref{eq:rew}. As our focus is on the relative behavior, we normalize each KPI to its maximum value. The effect of changing $\alpha_{\text d}$, i.e., the delay coefficient in the reward function is studied in Fig. \ref{figam}(a), where the other coefficients are set as $\alpha_{\text h}$=$\alpha_{\text f}$=0.5, and $\alpha_{\text s}$=0.01, i.e., we care more about delay and interference. One observes that by increasing  $\alpha_{\text d}$ from $0$ to $0.5$, which is   corresponding to the case that delay is less tolerable, more  handovers are triggered. This is due to the fact that the policy enforces handover to the BSs on which  drone experiences less path-loss in communications. Furthermore, it is clear that by connecting to the BS with the best channel, the level of interference will be decreased. Moreover, one observes that the proposed scheme significantly outperforms the benchmark at $\alpha_{\text d}$=$0.5$ in all KPIs. For $\alpha_{\text d}$ $\geq$ 0.5, we observe that the number of potential handover triggering epochs increases by 8\%, however, there is no negligible change in the performance of other KPIs, including the delay. The same results could be observed in Fig. \ref{figam}(b), where we  investigate the impact of $\alpha_{\text f}$, the interference coefficient in the reward function, on the policy design. One observes in this figure that by an increase in the  relative importance of   interference in the reward function, the number of handovers increases to avoid interference to the neighboring BSs. On the other hand, such an increase in $\alpha_{\text f}$ decreases the number of allocated resources to interfering drones. 

Fig. \ref{figam}(c) studies the effect of $\alpha_{\text h}$ as the coefficient of handover in \eqref{eq:rew}. As shown in this figure, an increase  in $\alpha_{\text h}$ from $0$ to $0.5$, i.e., an increase in the regret for handovers, decreases the potential decision epochs in which an unnecessary handover may occur  by $95\%$. Interestingly, one observes that this increase  causes higher regret for handover  has also improved the performance of all other KPIs. For example, one observes that the experienced delay has been decreased by 5-10\% because of the decrease in the control signaling for carrying handovers out.  On the other hand, for $\alpha_{\text h}$$>$0.5, we observe that almost all other KPIs, including experienced delay, buffer queue size, and level of incurred interference, are traded for further decrease in the number of handovers. 
 
\begin{figure}[!t]
    \centering
    \includegraphics[width=0.45\textwidth,height=2in]{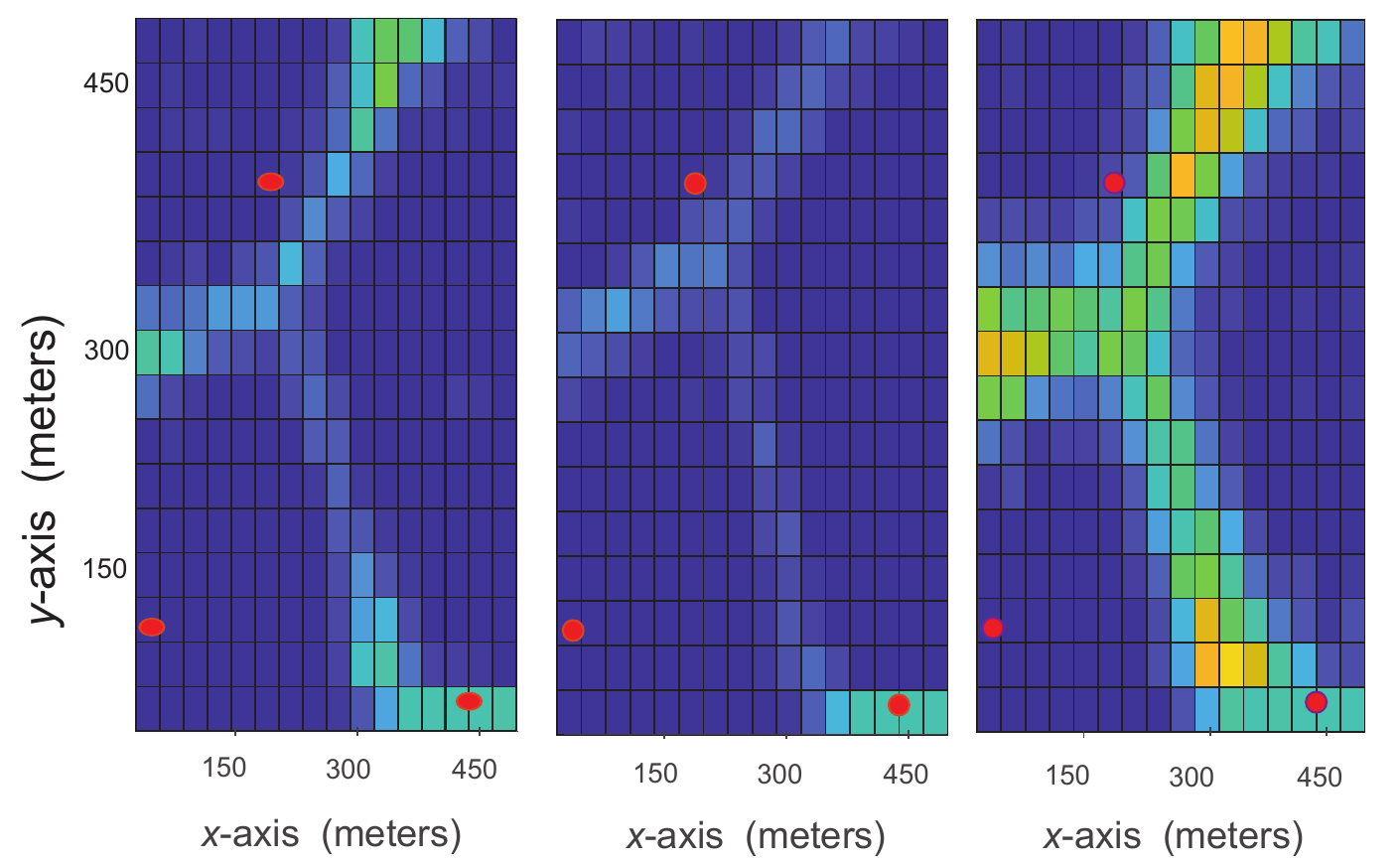}
     \caption{Heat-maps of handover positions. Red dots represent the BSs. Parameters:  $\alpha_{\text d}$=$\alpha_{\text f}$=$\alpha_{\text h}$=0.5. Left: $v=15$ m/s, $h=50$ m; Middle: $v$=60 m/s, $h$=50 m; Right: $v$=15 m/s, $h$=250 m. }
    \label{fig:heat1}
\end{figure}

The effects of altitude and velocity on handover have been  studied in Fig. \ref{fig:heat1}. This figure presents the heat-maps of positions at which, handover decisions have been made during offering service to $850$ drones. The locations of the BSs have been marked by red dots, drones are crossing the service area from left to right, and are initially to the BS at the bottom-left. Let us first focus on the impact of speed on handover decisions. The left and middle heat-maps correspond to $15$ m/s and $60$ m/s speeds, respectively. It is clear that by an increase in the speed, the handover decisions have been reduced significantly, and hence, we observe almost a cell-less connectivity in the sky. This is mainly due to the fact that while the interference, due to not triggering handover,  is suffering, but the time-length of such suffering is such low (due to the high speed), that the controller prefers not to trigger a handover if it is not really needed due to the delay requirement. Furthermore, the impact of altitude on the handovers can be seen by comparing the heat-map in left and right, where drone in the latter is $200$ m further away from the ground.  The increase in the frequency of handovers in the right heat-map could be reasoned by recalling equation \eqref{pl3}, in which we showed how  probability of experiencing LoS propagation increases by increasing the altitude. Then, cardinality of the set of BSs to which drone can handover in the right scenario is higher than the other cases, which results in further handovers for the drone. 

\section{Conclusion} 
This paper studies a learning-powered approach for handover and resource management in cellular networks serving drone users (the uplink direction). The major challenges consist of the interference from drones on uplink communications of coexisting terrestrial users, and the frequent handovers for drones.  The handover management and resource allocation optimization problem have been transformed to a machine learning problem for which a reinforcement learning algorithm is proposed as a solution. The design of this algorithm features incorporating different sources of rewards and regrets with respect to the network resources, and KPIs of drone users and interference to ground BSs. A comprehensive set of simulations have been conducted, where the results confirmed the significant impact of resource allocation and handover management for drone communications on terrestrial users. 
By setting appropriate coefficients for delay, interference and handover in the reward function, one can significantly  outperform the benchmark scheme in terms of number of handovers, incurred interference and experienced delay. Furthermore, an increase in the speed cause less handover decisions, however, altitude has the reverse effect on the number of handovers.


\bibliographystyle{IEEEtran}
\bibliography{IEEEabrv,Reference}

\end{document}